\documentclass[preprint,onecolumn,nofootinbib]{revtex4}
\pdfoutput=1
\usepackage[colorlinks=true,linkcolor=blue,urlcolor=blue,filecolor=black,citecolor=red,pdfstartview=FitV,pdftitle={},pdfsubject={},pdfkeywords={},pdfpagemode=None,bookmarksopen=true]{hyperref}
\usepackage{graphicx}%Include figure files
\usepackage{amsmath}
\usepackage{amsfonts}
\usepackage{amssymb,ulem}
\usepackage{dcolumn,booktabs}%
\usepackage{color,xcolor}%
\usepackage{CJK}
\usepackage{subfigure}

\usepackage{dcolumn}% Align table columns on decimal pointl
\usepackage{float}
\usepackage{multirow}% apply to table
\newcommand{\f}{\begin{equation}}
\newcommand{\ff}{\end{equation}}
\newcommand{\fa}{\begin{eqnarray}}
\newcommand{\ffa}{\end{eqnarray}}
\begin{document}
\title{Constraints from Solar System tests on a covariant loop quantum black hole}

\author{Ruo-Ting Chen$^{1}$}
\thanks{ruotingchen@163.com}
\author{Shulan Li$^{1}$}
\thanks{shulanli.yzu@gmail.com}
\author{Li-Gang Zhu$^{1}$}
\thanks{zlgoupao@163.com}
\author{Jian-Pin Wu$^{1}$}
\thanks{Corresponding author: jianpinwu@yzu.edu.cn} 
\affiliation{
  $^1$ Center for Gravitation and Cosmology, College of Physical
  Science and Technology, Yangzhou University, Yangzhou 225009,
  China}

\begin{abstract}

Recently, a covariant spherically symmetric model of a black hole within the framework of loop quantum gravity (LQG), characterized by a quantum parameter $r_0$ or $\lambda$, has been proposed. To derive constraints on the LQG-corrected parameter, we explore observational constraints imposed on $r_0$ and $\lambda$ through investigations of the light deflection, the Shapiro time delay, the precession of perihelia, and the geodetic precession test. Among these constraints, the tightest one arises from the Shapiro time delay measured by the Cassini mission, yielding an upper constraint of approximately $10^{-5}$.

\end{abstract}

\maketitle
\tableofcontents

\section{Introduction}

A variety of ground- and space-based precision experiments, including those related to the deflection of light, the Shapiro time delay, and the perihelion advance, have consistently validated the reliability of general relativity (GR) in the weak field regime \cite{Will:2014kxa}. Recent observations, such as the detection of gravitational waves (GWs) resulting from binary system mergers \cite{LIGOScientific:2016aoc,LIGOScientific:2016lio,LIGOScientific:2016sjg} and the imaging of supermassive black holes' shadows (M87$^*$ and $\mathrm{Sgr\ A^{*}}$) using the Event Horizon Telescope \cite{EventHorizonTelescope:2019dse,EventHorizonTelescope:2019ths,EventHorizonTelescope:2022xnr,EventHorizonTelescope:2022xqj}, not only confirm the existence of black holes but also serve as rigorous tests of the resilience of GR in the strong field regime.

However, despite these remarkable achievements, GR is still far from being a flawless theory. From a theoretical standpoint, developing a consistent quantum gravity theory that effectively reconciles GR and quantum mechanics remains the preeminent theoretical challenge in the field of fundamental physics. Among various approaches to quantum gravity, loop quantum gravity (LQG) distinguishes itself with its background independence, non-perturbative nature, and well-defined mathematical framework \cite{rovelli2004quantum, Ashtekar:2004eh}. It offers a promising avenue for understanding the quantum behavior of gravity.

Furthermore, by incorporating two key ingredients of LQG, namely the inverse volume correction and the holonomy correction, loop quantum cosmology (LQC) has been successfully formulated \cite{Bojowald:2001xe,Ashtekar:2006rx,Ashtekar:2006uz,Ashtekar:2006wn,Ashtekar:2003hd,Bojowald:2005epg,Ashtekar:2011ni,Wilson-Ewing:2016yan}. The quantum gravity effects in LQC can be linked to low-energy physics, offering a solvable cosmological model to explore quantum gravity phenomena. Interestingly, the quantum gravity effects in LQC successfully bypass the big bang singularity in classical GR \cite{Bojowald:2001xe,Ashtekar:2006rx,Ashtekar:2006uz,Ashtekar:2006wn,Ashtekar:2003hd,Bojowald:2005epg,Ashtekar:2011ni,Wilson-Ewing:2016yan,Bojowald:2003xf,Singh:2003au,Vereshchagin:2004uc,Date:2005nn,Date:2004fj,Goswami:2005fu, Papanikolaou:2023crz}, replacing it with a nonsingular big bounce even at the semiclassical level \cite{Bojowald:2005zk,Stachowiak:2006uh}.

Building upon the similar idea in LQC \cite{Bojowald:2001xe,Ashtekar:2006rx,Ashtekar:2006uz,Ashtekar:2006wn,Ashtekar:2003hd,Bojowald:2005epg,Ashtekar:2011ni,Wilson-Ewing:2016yan}, several effective black hole (BH) models incorporating LQG corrections have been developed. Notable examples of these models can be found in  \cite{Ashtekar:2005qt,Modesto:2005zm,Modesto:2008im,Modesto:2009ve,Campiglia:2007pr,Bojowald:2016itl,Boehmer:2007ket,Chiou:2008nm,Chiou:2008eg,Joe:2014tca,Yang:2022btw,Lewandowski:2022zce,Gan:2022oiy, Vagnozzi:2022moj, Afrin:2022ztr}, along with relevant references. The replacement of the singularity by a transition surface that connects a trapped region to an antitrapped region, which can be viewed as the inner region of a black hole and a white hole, is a typical feature of LQG-BHs.

Currently, the majority of effective LQG-BHs are implemented using the holonomy correction as an input. 
The phase space regularization technique known as polymerization is at the heart of the holonomy correction \cite{Corichi:2007tf}. As a result, the polymer BHs are another name for the effective LQG-BHs with holonomy correction. The basic idea underlying polymerization involves the substitution of the conjugate momentum $p$ with its regularized counterpart $\sin(\lambda p)/\lambda$, where $\lambda$ represents the polymerization scale, a parameter associated with the area-gap. 

In recent studies conducted by Alonso-Bardaji \textit{et al}. \cite{Alonso-Bardaji:2021yls, Alonso-Bardaji:2022ear}, a covariant model of a spherically symmetric black hole with holonomy correction is introduced, building upon the concept of anomaly-free polymerization as discussed in their previous work \cite{Alonso-Bardaji:2021tvy}. The quantum gravity effects are controlled by a quantum parameter $r_0$, which is a combination of the polymerization parameter $\lambda$ and the constant of motion $M$. This results in the formation of an interior region that is free from singularities, as well as two outer regions that approach flatness as they extend toward infinity. Notably, both outer regions possess the same mass. 

Subsequently, this LQG black hole solution has been extended by the authors to include charge in the cosmological background \cite{Alonso-Bardaji:2023niu}. Additionally, the authors have also explored this LQG model that coupled to matter \cite{Alonso-Bardaji:2021tvy, Alonso-Bardaji:2023vtl}. Furthermore, several investigations have already explored various aspects of this model. For example, the study of quasinormal modes (QNMs) of this LQG black hole has been carried out in \cite{Fu:2023drp, Moreira:2023cxy, Bolokhov:2023bwm}; the feasibility of this model extension to the Planck scale and a remnant one has been studied in \cite{Sobrinho:2022zrp, Borges:2023fub}; and gravitational lensing and optical behaviors have also been discussed in \cite{Soares:2023uup, Junior:2023xgl, Balali:2023ccr}. 

This work aims to investigate the classical tests of the covariant LQG-corrected black hole in the context of the solar system. These tests encompass the light deflection, the Shapiro time delay, the perihelion precession, and the geodetic precession. Classical detection methods within the solar system have been employed in numerous modified gravity models, such as those discussed in \cite{Farrugia:2020fcu, deng2017gravitational, Deng:2017hkj, Okcu:2021oke}, and even within the context of five-dimensional Kaluza-Klein gravity spacetime \cite{Liu:2000zq, Deng:2015sua}. Significantly, these classical detection methods in the solar system have been employed in recent studies to impose constraints on the LQG-corrected black hole \cite{Zhu:2020tcf, Liu:2022qiz}.
In this study, we examine how quantum gravity effects modify classical tests of GR predictions based on the behavior of test particles within the framework of covariant LQG black hole spacetime.
In each case, we perform a thorough analysis of our findings by utilizing high-precision datasets from solar system astronomical observations. Through this process, we derive numerical upper boundaries for the quantum parameters $r_0$ and $\lambda$. It is imperative to acknowledge that the primary emphasis of this article is only on the static spacetime, with the deliberate omission of the rotational influence commonly referred to as the Lense-Thirring effect \cite{park2017precession, Lucchesi:2010zzb}.

This paper is structured as follows. In Sec. \ref{bhsolution}, a concise overview of the covariant LQG black hole model is presented, along with an examination of the geodesic motion of a test particle within the framework of this LQG-corrected black hole. In Sec. \ref{constrain}, the modified formulas of classical tests of GR predictions incorporating LQG corrections are introduced. The obtained results are subsequently compared with the latest observational data from the solar system, leading to numerical constraints on the quantum parameters $r_0$ and $\lambda$. Finally, our findings and a brief outlook for potential advancements are summarized in Sec. \ref{conclusion}.

Throughout this paper we adopt Planck units, i.e. setting $G=c=\hbar=1$ in theoretical calculations, and utilize the $\left(-, +, +, +\right)$ signature for the metric. When comparing with data from the solar system, we revert to the international system of units. Latin letters represent abstract index notation, while Greek indices range over $0, 1, 2, 3$. We use Schwarzschild coordinate system $x^{\mu}=\left(x^{0}, x^{1}, x^{2} , x^{3}\right)\equiv \left(t, r, \theta, \phi\right)$.

%%%%%%%%%%%%%%%%%%%%%%%%%%%%%%%%%%%%%%%%%%
\section{Motion of a test particle over an effective covariant LQG black hole}\label{bhsolution}
In this section, we begin by providing a concise overview of the novel effective LQG black hole model, which incorporates holonomy corrections parametrized by a quantum parameter $r_0$ or $\lambda$. And then, we derive the equations of motion (EOM) for a test particle orbiting the black hole using the Hamiltonian canonical method.

\subsection{LQG black hole spacetime}

The spherically symmetric exterior geometry of this effective LQG black hole is described
as follows \cite{Alonso-Bardaji:2021yls, Alonso-Bardaji:2022ear}:
\begin{eqnarray}
\mathrm{d}s^{2}=-f\left(r\right) \mathrm{d}t^{2}+\frac{1}{g\left(r\right) f\left(r\right)} \mathrm{d}r^{2}+r^2 \left(\mathrm{d}\theta^2 +\sin^2\theta\thinspace \mathrm{d}\phi^2 \right) , \label{equ:II.A-(1)}
\end{eqnarray}
\begin{eqnarray}
f\left(r\right)=1-\frac{2 M}{r}, \quad g\left(r\right)=1-\frac{r_{0}}{r}.
\end{eqnarray}
A new length scale $r_0$ is introduced as a result of quantum gravity effects:
\begin{eqnarray}
r_{0}=2 M \frac{{\lambda}^{2}}{1+{\lambda}^{2}} .
\end{eqnarray}
Here, $\lambda$ is a dimensionless parameter inspired by holonomies, and without loss of generality, we can assume that $\lambda>0$. It is evident that the quantum parameter $r_0$ defines a minimum area gap ${r_0}^2$. $M$ represents the constant of motion, which is associated with the ADM mass as:
\begin{eqnarray}
\tilde{M} \equiv M_{\mathrm{ADM}} = M+\frac{r_0}{2} .
\end{eqnarray}
The ADM mass will also be identified as the celestial mass.
In the limit $\lambda \to 0$, yielding $r_0=0$, the effective LQG geometry described by Eq. \eqref{equ:II.A-(1)} regresses to the conventional Schwarzschild geometry of GR. 

Before proceeding, we would like to provide some insights into the interior geometry of this effective LQG black hole. Upon incorporating the LQG correction, the interior classical singularity is resolved by a minimal spacelike hypersurface at $r = r_0$, resulting in a connected region between a black hole and a white hole. This region is characterized by two external asymptotically flat areas of equal mass. In this scenario, the two-sphere bounce surface characterized by a minimal area of $4\pi r_0^2$ always hide inside the event horizon, i.e. $r_0<2M$. This region is also referred to as the black bounce over the global spacetime structure \cite{Moreira:2023cxy}. Notice that as the limit $M \to 0$ is approached, resulting in $r_0 \to 0$, the spacetime geometry reduces to a Minkowski configuration for any value of $\lambda$.

For the sake of convenient calculations throughout the paper, we will express the components of the metric \eqref{equ:II.A-(1)} in terms of the ADM mass $\tilde{M}$ and the dimensionless parameter $\tilde{r}_0$, which is defined as $\tilde{r}_0 \equiv {r_0}/\tilde{M}={2\lambda^2}/\left(1+2\lambda^2\right)$:
\begin{eqnarray}
f\left(r\right)=1-\frac{2{\tilde{M}}}{r}\left(1-\frac{\tilde{r}_0}{2}\right), \quad 
g\left(r\right)=1-\frac{{\tilde{M}}{\tilde{r}_0}}{r}.
\end{eqnarray}

To be able to reduce to the Newtonian limit, the gravitational constant in this effective LQG black hole can be related to the Newtonian gravitational constant by 
\cite{will2015gravity}
\begin{eqnarray}
	{G}_\mathrm{N} = G\left(1-\frac{\tilde{r}_0}{2}\right).
\end{eqnarray}
From now on, we will set ${G}_\mathrm{N}=1$ instead of $G=1$, which is more convenient for subsequent calculations.
Furthermore, for the remainder of the paper, we will eliminate the use of the tilde for the sake of simplicity.

%%%%%%%%%%%%%%%%%%%%%%%%%%%%%%
\subsection{Equations of motion for a test particle}

We commence by considering the Lagrangian governing the motion of a test particle over
the effective LQG black hole spacetime:
\begin{eqnarray}
\mathcal{L}=\frac{1}{2} m g_{\mu \nu} \dot{x}^{\mu} \dot{x}^{\nu}=\frac{1}{2} m g_{\mu \nu} \frac{\mathrm{d} x^{\mu}}{\mathrm{d} \tau} \frac{\mathrm{d} x^{\nu}}{\mathrm{d} \tau}\ .
\end{eqnarray}
Here, $m$ is the mass of the test particle, while $\tau$ can be chosen as the proper time or affine parameter for massive or massless particles along geodesics, with the overdot indicating derivative with respect to $\tau$.
Then the canonical momentum of the particle can be worked out as:
\begin{eqnarray}
p_{\mu } =\frac{\partial \mathcal{L} }{\partial \dot{x}^{\mu }  }= m g_{\mu \nu } \dot{x}^{\nu }  .
\label{equ:II.B-(7)}
\end{eqnarray}
Specially, we can explicitly express the four components of the canonical momentums as follows:
\begin{eqnarray}
&&p_{t} = m \left(-f\left(r\right)\right) \dot{t} , \label{equ:II.B-(8)} \\[3mm]
&&p_{r} = m \frac{1}{g\left(r\right) f\left(r\right)} \dot{r}, \\[3mm]
&&p_{\theta} = m r^{2} \dot{\theta}, \\[3mm]
&&p_{\phi} = m r^{2} \sin ^{2} \theta \thinspace \dot{\phi} . \label{equ:II.B-(11)}
\end{eqnarray}
%where $E$ indicates the conserved energy and $l$ indicates orbital angular momentum of the test particle.
Given that the Lagrangian does not depend on the variables $t$ and $\phi$, namely, ${\partial \mathcal{L}}/{\partial t}=0 $ and ${\partial \mathcal{L}}/{\partial \phi}=0 $, we have two Killing vectors, $\xi^{a} = \left({\partial}/{\partial t}\right)^{a}$ and $\eta^{a} = \left({\partial}/{\partial \phi}\right)^{a}$, which are associated with the energy ${E}$ and angular momentum ${l}$ of the test particle's motion, respectively. These quantities are determined by Eqs. \eqref{equ:II.B-(8)} and \eqref{equ:II.B-(11)}:
\begin{eqnarray}
{E} &=& {-p_{t}}\thinspace{\xi^{t}}  = m f\left(r\right) \dot{t} , \\[3mm]
{l} &=& {p_{\phi}}\thinspace{\eta^{\phi}} = m r^{2} \sin ^{2} \theta \thinspace \dot{\phi} .
\end{eqnarray}

By employing the Legendre transformation, we obtain the Hamiltonian $\mathcal{H}$ as follows:
\begin{eqnarray}
\mathcal{H} &=& p_{\mu} \dot{x}^{\mu }-\mathcal{L}
= \frac{1}{2m} \thinspace g^{\mu \nu} p_{\mu} p_{\nu}.
%&=&\frac{1}{2m}\left[-\frac{1}{f\left(r\right)} p_{t}^{2}+g\left(r\right) f\left(r\right) p_{r}^{2}+\frac{1}{r^{2}} p_{\theta}^{2}+\frac{1}{r^{2} \sin ^{2} \theta} p_{\phi}^{2}\right]
\end{eqnarray}
%\\[-5mm]
Then, we can explicitly derive the EOMs for the system. These equations are determined by evaluating the Poisson brackets between the canonical phase space variables and the Hamiltonian:
\begin{eqnarray}
&&\dot{t} =\left \{ t,\mathcal{H} \right \} =-\frac{1}{m f\left(r\right)}\thinspace p_{t} ,  \\[3mm]
&&\dot p_{t} =\left \{ p_{t},\mathcal{H} \right \} =0,  \label{equ:II.B-(20)} \\[3mm] %\quad \longrightarrow \quad p_{t}=-E=constant,  
&&\dot{r} =\left \{ r,\mathcal{H} \right \} = \frac{g\left(r\right) f\left(r\right)}{m}\thinspace p_{r} , \\[3mm]
&&\dot p_{r} =\left \{ p_{r},\mathcal{H} \right \} %=\frac{\partial p_{r}}{\partial r} \frac{\partial \mathcal{H}}{\partial p_{r}}-\frac{\partial p_{r}}{\partial p_{r}} \frac{\partial \mathcal{H}}{\partial r} = -\frac{\partial p_{r}}{\partial p_{r}} \frac{\partial \mathcal{H}}{\partial r} \nonumber \\[3mm]
%&& \qquad=-\frac{1}{2}\left(\frac{f^{\prime}\left(r\right)}{f^{2}\left(r\right)} p_{t}^{2}+g^{\prime}\left(r\right) f\left(r\right) p_{r}^{2}+f^{\prime}\left(r\right) g\left(r\right) p_{r}^{2}-\frac{2}{r^{3}} p_{\theta}^{2}-\frac{2}{r^{3}} \frac{1}{\sin ^{2} \theta} p_{\phi}^{2}\right) \nonumber \\[3mm]
=-\frac{f^{\prime}\left(r\right)}{2m f^{2}\left(r\right)}\thinspace p_{t}^{2}-\frac{g^{\prime}\left(r\right) f\left(r\right)+ g\left(r\right) f^{\prime}\left(r\right)}{2m}\thinspace p_{r}^{2}+ \frac{p_{\theta}^{2}}{m r^{3}} + \frac{p_{\phi}^{2}}{m r^{3} \sin ^{2} \theta},  \\[3mm]
&&\dot{\theta} =\left \{ \theta, \mathcal{H} \right \} =\frac{1}{m r^2}\thinspace p_{\theta},  \\[3mm]
&&\dot{p}_{\theta}= \left\{ p_{\theta}, \mathcal{H} \right \} %=\frac{\partial p_{\theta}}{\partial \theta} \frac{\partial \mathcal{H}}{\partial p_{\theta}}-\frac{\partial p_{\theta}}{\partial p_{\theta}} \frac{\partial \mathcal{H}}{\partial \theta}=-\frac{1}{r^{2}} \frac{-2}{\sin ^{3} \theta} \cos \theta \thinspace p_{\phi}^{2} \thinspace \frac{1}{2}
=\frac{\cos \theta}{m r^{2} \sin ^{3} \theta} \thinspace p_{\phi}^{2}, \\[3mm]
&&\dot{\phi}=\{\phi, \mathcal{H}\}=\frac{1}{m r^{2} \sin^{2} \theta} \thinspace p_{\phi}, \\[3mm]
&&\dot{p}_{\phi}=\left\{p_{\phi}, \mathcal{H} \right\}=0 . \label{equ:II.B-(26)}  %\quad \longrightarrow \quad p_{\phi}=l=constant. 
\end{eqnarray}
Equations \eqref{equ:II.B-(20)} and \eqref{equ:II.B-(26)} also indicate the conservation of energy and angular momentum for the test particle.

Utilizing these two constants of motion, we can express the reduced Hamiltonian as follows:
\begin{eqnarray}
\mathcal{H} = \frac{1}{2m}\left(-\frac{{E}^{2}}{f\left(r\right)} + g\left(r\right) f\left(r\right) \thinspace p_{r}^{2}+\frac{p_{\theta}^{2}}{r^{2}} +\frac{{l}^{2}}{r^{2} \sin ^{2} \theta} \right)
= \frac{\sigma}{2m} .
\label{equ:II.B-H}
\end{eqnarray}
%\\[-5mm]
In the above equation, we have utilized the normalization condition for four-velocity, denoted as $g^{\mu \nu} p_{\mu} p_{\nu} = {\sigma}$, where $\sigma$ takes different values depending on the nature of the particles involved. To be more specific, for massless particles, $\sigma=0$, while for massive particles, $\sigma=-m^2$.

%\begin{equation}
%-\frac{E^{2}}{f\left(r\right)}+g\left(r\right) f\left(r\right) \thinspace p_{r}^{2}+\frac{p_{\theta}^{2}}{r^{2}}+\frac{l^{2}}{r^{2} \sin ^{2} \theta}=  \sigma .
%\end{equation}

By employing the variable separation technique, we can identify a third constant of motion as follows:
\begin{eqnarray}
\mathcal{K} &=& p_{\theta}^{2}+\frac{l^{2}}{\sin ^{2} \theta} = r^{2} \thinspace \sigma+\frac{r^{2} E^{2}}{f\left(r\right)}-r^{2} g\left(r\right) f\left(r\right) \thinspace p_{r}^{2}.
\label{equ:II.B-(29)}
\end{eqnarray}
This separation constant $\mathcal{K}$ is commonly referred to as the Carter constant \cite{Carter:1968rr}. 

Next, we derive the equations for $r$ and $\theta$, which are expressed in terms of the aforementioned three constants of motion:
\begin{eqnarray}
&& \left(\frac{\mathrm{d} \theta}{\mathrm{d} \tau}\right)^{2}=\frac{\tilde{\mathcal{K}}}{\thinspace r^{4}}-\frac{\tilde{l}^{2}}{ r^{4} \sin ^{2} \theta }, \\[3mm]
&& \left(\frac{\mathrm{d} r}{\mathrm{d} \tau}\right)^{2} = -\frac{\tilde{\mathcal{K}} \thinspace g\left(r\right) f\left(r\right)}{ r^{2}} + \tilde{\sigma} \thinspace g\left(r\right) f\left(r\right) + \tilde{E}^{2} \thinspace g\left(r\right).
\end{eqnarray}
In the above calculations, we have introduced the variables per unit mass as: $\tilde{E} =E/m $, $\tilde{l} = l/m $, $\tilde{\mathcal{K}} = \mathcal{K}/m^2 $, $\tilde{\sigma} = \sigma/m^2 $. Once again, the tilde is also dropped for notational simplicity in the following paper. Thus, the geodesic equations for test particles then yield:
\begin{eqnarray}
\frac{\mathrm{d} t}{\mathrm{d} \tau} &=& \frac{{E}}{f\left(r\right)},   \label{equ:II.B-(32)}\\[3mm]
\left(\frac{\mathrm{d} r}{\mathrm{d} \tau}\right)^{2} &=& -\frac{{\mathcal{K}} g\left(r\right) f\left(r\right)}{r^{2}}+ \sigma g\left(r\right) f\left(r\right) + {E}^{2} g\left(r\right) ,   \label{equ:II.B-(33)} \\[3mm]
\left(\frac{\mathrm{d} \theta}{\mathrm{d} \tau}\right)^{2} &=& \frac{{\mathcal{K}}}{r^{4}}-\frac{{l}^{2}}{r^{4} \sin ^{2} \theta} ,   \label{equ:II.B-(34)} \\[3mm]
\frac{\mathrm{d} \phi}{\mathrm{d} \tau} &=& \frac{{l}}{r^{2} \sin ^{2} \theta} .  \label{equ:II.B-(35)}
\end{eqnarray}
Once we have the above equations available, we can delve deeper into exploring constraints on the quantum parameter through solar system experiments.

%%%%%%%%%%%%%%%%%%%%%%%%%%%%%%%%%%%%%%%%%%
\section{Constraints on quantum parameter}\label{constrain}

This section is dedicated to exploring the constraints placed on the quantum parameter through solar system experiments. These experiments include the deflection of starlight, the Shapiro time delay, the perihelion shift, and the geodetic precession.

\subsection{Deflection of light}\label{deflection}

Without loss of generality, we can solely focus on the evolution of motion is in the equatorial plane, where $\theta={\pi}/{2}$, then $\dot{\theta}=0$.
As a result, the Carter constant in Eq. \eqref{equ:II.B-(29)} simplifies to $\mathcal{K} = l^{2}$.

Let us consider a scenario where a light ray originates from infinity, gets deflected by the sun, and then escapes back to infinity. By utilizing Eqs. \eqref{equ:II.B-(33)} and \eqref{equ:II.B-(35)}, we can determine the trajectory of the light ray as follows:
\begin{eqnarray}
\frac{\mathrm{d} \phi}{\mathrm{d} r} = \pm\left(\frac{E^2 r^{4} g\left(r\right)}{l^{2}}-g\left(r\right) f\left(r\right) \thinspace r^{2}\right)^{-\frac{1}{2}} .
\label{equ:III.A-dr/dphi}
\end{eqnarray}
In this equation, the minus sign corresponds to photons moving inward with decreasing $r$, while the plus sign indicates outward-moving photons with increasing $r$. 
Subsequently, we will introduce the impact parameter $b$, which represents the perpendicular distance from the straight line of motion to the centerline of the sun that is parallel to it.
When the light ray at infinity, i.e. $r \to \infty$ and $\phi \to 0$, we get $\sin{\phi} \approx \phi = {b}/{r}$. By solving the Eq. \eqref{equ:III.A-dr/dphi} under this scenario, we approximately have:
\begin{eqnarray}
r = \pm \frac{l}{E \phi} .
\label{rphi-app}
\end{eqnarray}
It's easy to infer that $b=l/E$.
When the light ray approaches the sun and is affected by the gravitational field, it naturally reaches a turning point known as the closest approach. This point is located at a distance of $s$ from the center of the sun, at which the Eq. \eqref{equ:III.A-dr/dphi} vanishes; in other words, $\left({\mathrm{d} r}/{\mathrm{d} \phi}\right)_{r=s}=0$.
This leads to the following relationship:
\begin{equation}
b=\left( \frac{s^{2}}{f\left(s\right)}\right)^{\frac{1}{2}}.
\label{equ:III.A-b}
\end{equation}

Particularly, the magnitude of the total change in the coordinate interval $\phi$ is just twice the change in angle from the turning point $r=s$ to infinity. Therefore, we can express it in terms of $b$ as follows:
\begin{eqnarray}
\phi = 2 \int_{s}^{\infty}\left(\frac{r^{4} g\left(r\right)}{b^{2}}-g\left(r\right) f\left(r\right) r^{2}\right)^{-\frac{1}{2}} \mathrm{d} r.
\label{equ:III.A-phi}
\end{eqnarray}
In the absence of the sun, the light ray propagates in a straight line, where $\phi=\pi$. Consequently, the deflection angle $\Delta{\phi}$ is related to $\phi$ as: $\Delta \phi= \phi-\pi$.
%\begin{equation}
%\Delta \phi= \phi-\pi = 2 \int_{s}^{\infty}\left(\frac{r^{4} g\left(r\right)}{b^{2}}-g\left(r\right) f\left(r\right) r^{2}\right)^{-\frac{1}{2}} \mathrm{d} r-\pi.
%\label{equ:III.A-delta phi}
%\end{equation}
To proceed, we introduce a coordinate transformation $r={s}/{x}$ and define $\epsilon \equiv M/s$. And then, we can expand the equation of $\Delta{\phi}$ in the weak field approximation as follows:
\begin{eqnarray}
\Delta \phi=2 \int_{0}^{1} \left\{\frac{1}{\sqrt{1-x^{2}}}+\frac{\left[-r_0 + 2\left(1 + x + x^2 \right)\right] \epsilon }{\left(2-r_{0}\right) \left(1+x\right) \sqrt{1-x^{2}}} + \mathcal{O}\left(\epsilon^2\right) \right\} \mathrm{d}x - \pi . 
\label{equ:III.A-angle}
\end{eqnarray}

Therefore, we can obtain the following approximate expression for the light deflection angle with the quantum-corrected term:
\begin{eqnarray}
	\Delta \phi \approx % 4\epsilon + r_0 (\epsilon + \frac{3\pi{\epsilon}^2}{4}-{\epsilon}^2)
	\frac{4 M}{s}\left(1+\frac{r_{0}}{4 - 2 r_{0}}\right) 
	=\Delta \phi_{\rm{GR}}\left(1+\frac{r_{0}}{4 - 2 r_{0}}\right).
	\label{equ:III.A-f.o.-r0}
\end{eqnarray}
Here, $\Delta \phi_{\mathrm{GR}}$ represents the deflection value in GR. Notice that in the above calculation, we have reverted $\epsilon$ back to $M/s$.
 
When the light ray just grazes the sun, we assume that the closest approach $s$ is equal to the radius of the sun $R_{\odot}$, and $M$ is the solar mass $M_{\odot}$. In this scenario, the parameterized post-Newtonian (PPN) formalism equation for light deflection is given as follows \cite{will2018theory}:
\begin{eqnarray}
	\Delta \phi \simeq 1.75''\left(\frac{1+\gamma }{2}\right),  \label{equ:III.A-PPN}
\end{eqnarray}
where $\gamma$ is the PPN deflection parameter \cite{robertson1962space, eddington1923mathematical}. According to the astrometric observation measuring $\gamma$ by the Very Long Baseline Array (VLBA) \cite{Fomalont:2009zg}, we compare Eq. \eqref{equ:III.A-f.o.-r0} with Eq. \eqref{equ:III.A-PPN}. Consequently, the constraint on the quantum-corrected parameter $r_0$ can be immediately determined as follows:
\begin{eqnarray}
	0 <r_0 < 2.0\times 10^{-4}.
	\label{equ:III.A-L.D.-r0}
\end{eqnarray}
This leads to corresponding constraints on $\lambda$ where
\begin{eqnarray}
	0 < \lambda < 1.0\times 10^{-2}.
\end{eqnarray}

%%%%%%%%%%%%%%%%%%%%%%%%%%%%%%%%%%%%%%%%%%
\subsection{Shapiro time delay}

In this subsection, we will analyze the constraints on the LQG-corrected parameter through the study of the Shapiro time delay. Specially, we will consider a simplified scenario in which a radar signal is transmitted from a transmitter located on earth, denoted as $r=A$, then it passes through the closest approach to the sun at the turning point $r=s$, and finally returns to earth by reflection from a spacecraft-mounted reflector, denoted as $r=B$.

By combining Eqs. \eqref{equ:II.B-(32)} and \eqref{equ:II.B-(33)}, we can derive the differential equation for massless particles between $t$ and $r$:
\begin{eqnarray}
\frac{\mathrm{d}t}{\mathrm{d}r} 
=\pm \frac{r}{f\left(r\right)\sqrt{g\left(r\right)\left(r^2-b^2f\left(r\right)\right)} }  ,
\label{equ:III.B-dt/dr}
\end{eqnarray}
where the plus sign and the minus sign correspond to the outgoing and incoming radar waves, respectively.

Then, we can determine the travel time of the radar wave propagating from the transmitter at point $A$ to the turning point at point $s$:
\begin{eqnarray}
\Delta t_A = -\int_{A}^{s}\frac{r}{f\left(r\right)\sqrt{g\left(r\right)\left(r^2-b^2f\left(r\right)\right)} }\mathrm{d}r.
%&=& \int_{s}^{A}\frac{r}{f\left(r\right)\sqrt{g\left(r\right)(r^2-b^2f\left(r\right))} }dr \thinspace + \thinspace\int_{s}^{B}\frac{r}{f\left(r\right)\sqrt{g\left(r\right)(r^2-b^2f\left(r\right))} }dr .   
\label{equ:III.B-delta tA}
\end{eqnarray}
To evaluate the integral $\Delta t_A$, we once again use the coordinate transformation $r=s/x$ and expand the integral using the small quantity $\epsilon \equiv M/s$.
%%%
%\begin{eqnarray}
%\Delta t_A = \int_{\frac{s}{A} }^{1} \left( \frac{M}{x^{2} \epsilon \sqrt{1-x^{2}}} + \frac{M\left(4 - r_0 + 6x - 2 r_0 x \right)}{\left(2-r_0\right) x \left(1+x\right) \sqrt{1-x^{2}}} 
%%&+& \left. \frac{M\left(3 r_0^{2} \left(1+x\right)^{2}+4 r_0\left(2+5 x+3 x^{2}\right)+12\left(4+8 x+5 x^{2}\right)\right) \epsilon}{8\left(1+x\right)^{2} \sqrt{1-x^{2}}}
%+\mathcal{O}\left(\epsilon \right)\right)\mathrm{d}x .
%\end{eqnarray}
By integrating to the subleading order, we obtain:
\begin{eqnarray}
\Delta t_A &\approx& \sqrt{A^2-s^2} + M \sqrt{\frac{A-s}{A+s} } + M \left(\frac{4-r_0}{2-r_0}\right)\tanh^{-1}\left(\sqrt{1-\frac{s^2}{A^2} } \right) .
%&-& \frac{1}{2} m \epsilon \left(\frac{\sqrt{A-s}(A(4+r_0)+(5+r_0) s)}{(A+s)^{3 / 2}}-6(5+r_0) arcsin\left[\frac{\sqrt{1-\frac{s}{A}}}{\sqrt{2}}\right]\right) .
\end{eqnarray}
Similarly, the time it takes for the radar wave to travel between the turning point at $s$ and the reflector at $B$ can be determined in the same manner, and we will refer to it as $\Delta t_B$.

Based on the position of the spacecraft carrying the reflector, we usually classify it into two scenarios: inferior conjunction and superior conjunction when calculating the gravitational time delay. In the case of inferior conjunction, the spacecraft is situated between the earth and the sun. In the absence of gravitational effects, the total roundtrip time of the radar signal can be expressed as $\Delta t_{\mathrm{I-S}} = 2\left(\sqrt{A^2-s^2} -\sqrt{B^2-s^2}\right)$. When considering gravitational effects with LQG corrections, the roundtrip time delay is calculated as follows:
\begin{eqnarray}
\Delta t_{\mathrm{I}-r_0} 
\approx 4 M \left(\frac{4-r_0}{4 - 2 r_0}\right) \ln \left( \frac{A}{B}\right)
= \Delta t_{\mathrm{I-GR}}\left(\frac{4-r_0}{4 - 2 r_0}\right) ,
\end{eqnarray}
where $\Delta t_{\rm{I-GR}}$ represents the Shapiro time delay in GR. It is evident that the roundtrip time delay receives the LQG corrections.

In the case of superior conjunction, the spacecraft is located at the opposite side of the earth with respect to the sun. Therefore, in the absence of gravitational effects, the total roundtrip time for the radar wave to travel is $\Delta t_{\mathrm{S-S}} = 2\left(\sqrt{A^{2}-s^{2}} + \sqrt{B^{2}-s^{2}}\right)$.
Then, the roundtrip time delay with LQG corrections is calculated as follows:
\begin{eqnarray}
\Delta t_{\mathrm{S}-r_0}
\approx 4M \left[1 + \left(1+\frac{r_{0}}{4 - 2 r_{0}}\right)\ln\left(\frac{4AB}{s^{2}}\right)\right].
\label{equ:III.B-t-r0} 
\end{eqnarray}
In fact, the above equation is reformulated in the PPN-like formalism of the Shapiro time delay, which is \cite{will2018theory, Weinberg:1972kfs}:
\begin{eqnarray}
	\Delta t \simeq 4M \left[1 + \left(\frac{1+\gamma}{2}\right) \ln\left(\frac{4AB}{s^{2}}\right)\right].
\end{eqnarray}
It is equivalent to the relation between $r_0$ and $\gamma$ given in Sec. \ref{deflection}.

Next, we will use the Cassini solar conjunction mission in 2002 \cite{iess2003cassini, bertotti2003test} to constrain LQG-corrected parameter. By utilizing a multifrequency link in the X and Ka bands to minimize the influence of solar corona noise, significant improvements have been achieved, resulting in $\gamma=1+\left(2.1 \pm 2.3 \right)\times 10^{-5}$ \cite{bertotti2003test, iess1999doppler, Deng:2015sua, Deng:2017hkj}. Then we can give rise to the upper constraint on $r_0$ as:
\begin{eqnarray}
      0 <r_0 < 8.80\times 10^{-5} . 
\end{eqnarray}
The corresponding constraint on $\lambda$ is as follows:
\begin{eqnarray}
	 0 < \lambda < 6.63\times 10^{-3}.
\end{eqnarray}

On the other hand, we can also constrain the LQG parameter using the Doppler tracking of the Cassini spacecraft \cite{bertotti1992relativistic, bertotti1993doppler}. In contrast to the Shapiro time delay, which measures the time delay, the Doppler tracking directly measures the relative frequency variation. To achieve this, we differentiate Eq. \eqref{equ:III.B-t-r0} with respect to time $t$, leading to the fractional frequency shift for the radar signal \cite{iess1999doppler, Deng:2015sua}:
\begin{eqnarray}
	\frac{\mathrm{d} \Delta t_{\mathrm{S}-r_0}}{\mathrm{d} t} \equiv \delta \nu = \frac{\nu\left ( t \right )-\nu_0 }{\nu_0}
	\approx \left[-\frac{8 M}{s}-\frac{4 M r_0}{\left(2-r_0\right)s}\right]\frac{\mathrm{d} s(t)}{\mathrm{d} t}
	\approx \delta \nu_{\mathrm{GR}}-\frac{4 M r_0}{\left(2-r_0\right)s}\nu_\oplus ,
\end{eqnarray}
where $\nu_0$ and $\nu(t)$ are the emitted and received frequencies, respectively, and ${\mathrm{d} s(t)}/{\mathrm{d} t}$ is approximately equivalent to the average orbit velocity of earth, denoted as $\nu_\oplus$. Therefore, the frequency shift caused by the quantum correction parameter $r_0$ can be expressed as \cite{Deng:2017hkj, Zhu:2020tcf, Liu:2022qiz}:
\begin{eqnarray}
	\left | \delta \nu_{\mathrm{S}-r_0} \right |  \approx \frac{4 M r_0}{\left(2-r_0\right)s}\nu_\oplus = \frac{4 M_\odot r_0}{\left(2-r_0\right) R_\odot} \frac{16}{27}\nu_\oplus < 10^{-14} ,
\end{eqnarray}
where $M_\odot$ and $R_\odot$ respectively denote the mass and radius of the sun. 
By taking these conditions into account, we obtain an upper constraint on $r_0$ within the range of $0 <r_0 < 4.0 \times 10^{-5}$, along with corresponding constraints on $\lambda$ falling in the range of $0 < \lambda < 4.47 \times 10^{-3}$. It is evident that both the Shapiro time delay and the Doppler tracking of the Cassini spacecraft yield consistent results.

%%%%%%%%%%%%%%%%%%%%%%%%%%%%%%%%%%%%%%%%%%
\subsection{Precession of perihelia}

In this subsection, we investigate the constraints on the LQG-corrected parameter through the study of perihelion precession. To do so, we analyze the motion of a massive particle $(\sigma=-1)$ orbiting the sun. 

First, by combining Eq. \eqref{equ:II.B-(33)} and Eq. \eqref{equ:II.B-(35)}, we can derive the equation describing the orbit precession of the massive particle as follows:
\begin{eqnarray}
\left(\frac{\mathrm{d}x}{\mathrm{d}\phi}\right)^{2} = \left[\frac{s^{2}}{l^{2}}\left(1-f\left(\frac{s}{x}\right)\right) + \frac{s^{2}}{b^{2}} - f\left(\frac{s}{x}\right) x^2 \right] g\left(\frac{s}{x}\right) ,
\end{eqnarray}
where we have also introduced the coordinate transformation $r=s/x$ as previously utilized in Sec. \ref{deflection}. In addition, the impact parameter relates both $l$ and $E$ as $b=l/\sqrt{E^2-1}$ in the  case of massive particle.

When we differentiate the equation above with respect to $\phi$ and then expand the formula within the weak field limit, defined in terms of the small parameter $\epsilon$ where $\epsilon \equiv M/s$, we obtain an approximate description of the revolution of the orbits, given by:
\begin{eqnarray}
\frac{\mathrm{d}^2 x}{\mathrm{d} \phi^2}+x-\frac{M^2}{l^{2}\epsilon} \approx 3 \epsilon x^{2} + \frac{r_0\left[b^2 \epsilon^2 x^{2}l^2\left(3-8\epsilon x\right) - M^2\left(l^2 + 4 b^2 \epsilon x\right)\right]}{ b^2 l^{2} \left(2-r_0\right) \epsilon }.
\label{equ:III.C-D.E.}
\end{eqnarray}
We can clearly observe that the LQG effect is distinctly manifested in the second term on the right-hand side of Eq. \eqref{equ:III.C-D.E.}.
In what follows, we will solve the above differential equation using the perturbation methods.

To do this, we begin by expressing $x\left(\phi \right)$ as $x\left(\phi \right)=x_0\left(\phi \right)+x_1\left(\phi \right)$, with the condition that $x_1\left(\phi \right) \ll x_0\left(\phi \right)$.
When the left-hand side of Eq. \eqref{equ:III.C-D.E.} equals zero, the situation reverts to the Newtonian gravity theory. In this case, the solution reads
\begin{eqnarray}
x_0\left(\phi \right)=\frac{M^2}{l^2 \epsilon} \left(1+e \cos\phi \right) ,
\label{equ:III.C-newton}
\end{eqnarray}
which is the unperturbed part and is commonly known as the conic section formula with eccentricity $e$ involved.

Next, we will determine the perturbation part $x_1\left(\phi \right)$. To this end, we substitute the expression $x\left(\phi \right)=x_0\left(\phi \right)+x_1\left(\phi \right)$, with $x_0\left(x\right)$ obtained in Eq. \eqref{equ:III.C-newton}, into Eq. \eqref{equ:III.C-D.E.}, while considering the initial conditions $x_1\left(0\right)=0$, ${\mathrm{d}x_1\left(0\right)}/{\mathrm{d}\phi}=0$. This yields the following equation:
\begin{eqnarray}
\frac{\mathrm{d}^2 x_1}{\mathrm{d} \phi^2} + x_1 
= \sum_{i=0}^{3} \chi_{i} \cos^{i}\phi ,
\label{equ:III.C-x1-D}
\end{eqnarray}
where
%\begin{eqnarray}
%\chi_{0} &=& \frac{\left(1-E^{2}\right) l^{4} M^{2} r_0 - 8 M^{6} r_0 + 2l^{2} M^{4}\left(3-2 r_0\right)}{l^{6}\left(2-r_0\right) \epsilon} , \\[3mm]
%\chi_{1} &=& \frac{4 M^{4} e\left(l^{2}\left(3-r_0\right)-6 M^{2} r_0\right)}{l^{6}\left(2-r_0\right) \epsilon} , \\[3mm]
%\chi_{2} &=& \frac{6M^{4} e^{2}\left(l^{2}-4 M^{2} r_0\right)}{l^{6}\left(2-R\right)\epsilon} , \\[3mm]
%\chi_{3} &=& -\frac{8 M^{6} e^{3} r_0}{l^{6}\left (2-r_0\right ) \epsilon} .
%\end{eqnarray}
\begin{eqnarray}
\chi_{0} &=& \frac{b^{2}\left[-8 M^{6} r_0+2l^{2} M^{4}(3-2 r_0)\right]-l^{6} M^{2} r_0}{b^{2} l^{6} (2-r_0)\epsilon} , \\[3mm]
\chi_{1} &=& \frac{4 M^{4} e\left[l^{2}(3-r_0)-6 M^{2} r_0\right]}{l^{6}(2-r_0)\epsilon} , \\[3mm]
\chi_{2} &=& \frac{6 M^{4} e^{2}\left(l^{2}-4 M^{2} r_0\right)}{l^{6}(2-r_0)\epsilon} , \\[3mm]
\chi_{3} &=& -\frac{8 M^{6} e^{3} r_0}{l^{6}(2-r_0)\epsilon} .
\end{eqnarray}
\\[-5mm]
We can therefore obtain the solution:
\begin{eqnarray}
x_1(\phi) &=& \chi_{0} + \frac{\chi_{2}}{2}-\chi_{0}\cos\phi-\frac{\chi_{2}}{3}\cos\phi+\frac{\chi_{3}}{32}\cos\phi-\frac{\chi_{2}}{6}\cos\left(2\phi\right)-\frac{\chi_{3}}{32}\cos\left(3\phi\right) \nonumber \\[3mm]
&+& \frac{\chi_{1}}{2}\phi \sin\phi + \frac{3 \chi_{3}}{8}\phi \sin\phi .
\label{equ:III.C-x1-S}
\end{eqnarray}
Regarding the perihelion precession, when the terms involving $\phi\sin\phi$ in Eq. \eqref{equ:III.C-x1-S} are absent, the test particle remains on a closed orbit without deflection. As time progresses, the cumulative effect makes the perihelion precession of planetary orbits observable. Therefore, in this scenario, the remaining terms in Eq. \eqref{equ:III.C-x1-S} can be omitted.
Finally, we obtain the approximate solution to Eq. \eqref{equ:III.C-D.E.} as follows:
\begin{eqnarray}
 x\left(\phi\right) \approx \frac{M s}{l^2} \left(1 + e \cos\phi \right) + \left( \frac{\chi_{1}}{2} + \frac{3 \chi_{3}}{8} \right) \phi \sin\phi
\approx \frac{M s}{l^2} \left(1 + e \cos \left(\phi - \phi_0 \right)\right) ,
\end{eqnarray}
where we have transformed $\epsilon$ back into $M/s$, and now we can relate the precession angle $\delta\phi_0$ using the expression $\phi_0 = \left(\delta \phi_0/2 \pi \right)\phi$ as follows:
\begin{eqnarray}
\delta{\phi_0} \approx \frac{4\pi M^{2}}{l^{2}}\left(\frac{3-r_0}{2-r_0}\right) .
\label{equ:III.C-phi0}
\end{eqnarray}
In particular, the radial distance $r$ attains its minimum value at perihelia, where the condition $\phi - \phi_0 = 0$ yields the equation $s/r_{-}={Ms\left(1+e\right)}/{l^2}$. On the other hand, the radial distance achieves its maximum value at aphelia, where the condition $\phi - \phi_0 = \pi$ results in the equation $s/r_{+} = {Ms\left(1-e\right)}/{l^2}$. While for any bound orbit, we can determine the semimajor axis $a$ using the following formula:
\begin{eqnarray}
a = \frac{r_{-} + r_{+}}{2}=\frac{l^2}{M(1-e^2)} .  
\label{equ:III.C-a}
\end{eqnarray}
Combining Eqs. \eqref{equ:III.C-phi0} and \eqref{equ:III.C-a}, we obtain the angle of perihelion precession per revolution deviated from the GR prediction:
\begin{eqnarray}
	\Delta{\phi} = \delta{\phi_0} \approx \frac{6 \pi M}{a \left(1-e^2 \right)}\left(\frac{6-2r_0}{6-3r_0} \right)  = \Delta{\phi}_{\rm{GR}} \left(1+\frac{r_0}{6-3r_0} \right) .
\label{equ:III.C-delta phi0}
\end{eqnarray}
Additionally, we can express the $\Delta{\phi}_{\rm{GR}}$ in terms of the solar mass $M_{\odot}$ as $\Delta{\phi}_{\rm{GR}} = {6 \pi M_{\odot }}/{\left[ a\left(1-e^2 \right)\right]}$.

For the observation of Mercury's anomalous perihelion advance, the MESSENGER mission provided highly accurate measurements \cite{park2017precession}, yielding a value of $\Delta{\phi} = \left(42.9799 \pm 0.0009\right)''$ per century. Using this measured data, we can establish upper bounds on the parameters $r_0$ and $\lambda$, resulting in the following constraints:
\begin{eqnarray}
0 <r_0 < 1.26\times 10^{-4} , \quad  0 < \lambda < 7.93\times 10^{-3}.
\end{eqnarray}
This result aligns with the expectations that the contribution from LQG effect is less than the observational error $0.0009''$ per century. In Appendix \ref{Appendix}, we provide further discussions on this topic using the PPN method.

It should be noted that Eq. \eqref{equ:III.C-delta phi0} only considers the gravitoelectric perihelion shift  resulting from the influence of the solar mass $M_{\odot }$, whereas the gravitomagnetic component, commonly referred to as the Lense-Thirring effect, is not taken into account within in this equation \cite{lense1918einfluss}. Based on the summary provided in \cite{park2017precession}, it is observed that the level of uncertainty associated with the total precession rate is comparatively smaller than the estimated contributions attributed to the Lense-Thirring effect over a century. Consequently, the measurement of the Lense-Thirring effect has not yet attained a commensurate level of precision. As a result, for the purposes of this paper, it is deemed appropriate to disregard this effect and consider the central object as nonrotating.

Hence, the LAGEOS satellites are taken into consideration due to their ability to yield precise outcomes through the measurement of the relativistic precession of LAGEOS II's pericenter within the earth's orbit \cite{Lucchesi:2010zzb}. Based on the analysis of tracking data spanning a period of 13 years, the PPN level factor $\epsilon_{\omega}$ is estimated to be $\epsilon_{\omega}=1+\left(0.28 \pm 2.14\right)\times 10^{-3}$.  When $\epsilon_{\omega}=1$, the situation reverts back to the case of GR. When compared to the excessive coefficient in Eq. \eqref{equ:III.C-delta phi0}, it results in the constraints on $r_0$ and $\lambda$ as follows:
\begin{eqnarray}
0 <r_0 < 1.44\times 10^{-2} ,\quad  0 < \lambda < 8.55\times 10^{-2}.
\end{eqnarray}

Additionally, the observations of star S2 orbit around $\mathrm{Sgr\ A^{*}}$, the closest massive black hole candidate at the centre of the Milky Way, provide an alternative means of testing the Schwarzchild precession (SP) \cite{GRAVITY:2020gka}.
GR predicts a precession advance angle of $\Delta {\phi}_{\rm{GR}} = 12.1'$ per orbital period.
The data analysis by the GRAVITY collaboration yields a PPN-like parameter $f_{\rm{SP}}$. In the context of GR, it is anticipated that this parameter would have a value of $1$. Nevertheless, a fiducial value with uncertainty $f_{\rm{SP}} = 1.10 \pm 0.19$ was determined. Comparing these results with Eq. \eqref{equ:III.C-delta phi0}, the corresponding upper bounds on $r_0$ and $\lambda$ are as follows:
\begin{eqnarray}
0 <r_0 < 0.93 , \quad  0 < \lambda < 2.59.
\end{eqnarray}
This is in agree with the result given in \cite{Balali:2023ccr}, where $\lambda \in \left [ 0, 2.65 \right ] $ is obtained by using the $\mathrm{Sgr\ A^{*}}$ shadow's angular diameter.

%%%%%%%%%%%%%%%%%%%%%%%%%%%%%%%%%%%%%%%%%%
\subsection{Geodetic precession}

In this subsection, our attention turns to another test of GR known as geodetic precession. This test serves the purpose of probing the spacetime geometry and place constraints on the LQG parameter.

We commence by studying the motion of the spin of a point test particle in free fall \cite{Schiff:1960gi, hartle2003gravity}.
We assume that the test particle moves along a timelike geodesic, whose four-velocity vectors $u^{\mu}=\dot{x}^{\mu}={\mathrm{d} x^{\mu}}/{\mathrm{d}\tau}$, governed by the geodesic equation:
\begin{eqnarray}
\frac{\mathrm{d} u^{\mu}}{\mathrm{d} \tau}+\Gamma _{\ \nu \alpha}^{\mu} u^{\nu} u^{\alpha} &=& 0 , \label{equ:III.D-geodesic} 
\end{eqnarray}
where $\Gamma _{\enspace \nu \alpha}^{\mu}$ represents the four-dimensional Christoffel symbol.
The evolution of its spin four-vector $S^{\mu}$ along the geodesic is described as follows:
\begin{eqnarray}
\frac{\mathrm{d} S^{\mu}}{\mathrm{d} \tau}+\Gamma _{\ \nu \alpha}^{\mu} S^{\nu} u^{\alpha} = 0 . \label{equ:III.D-parallel}
\end{eqnarray}
This equation is commonly referred to as the gyroscope equation or parallel transport equation \cite{walker1935note, synge1960relativity, Misner:1973prb}. 
In addition, we will use the orthogonality and normalization conditions, which are expressed as:
\begin{eqnarray}
u^{\mu} S_{\mu}&=&0 ,  \label{equ:III.D-orth} \\[1mm]
S^{\mu} S_{\mu}&=&1 .  \label{equ:III.D-norm} 
\end{eqnarray}
%\begin{eqnarray}
%\Gamma _{\enspace \nu \alpha}^{\mu}=\frac{1}{2}g^{\mu \rho}\left(g_{\rho \nu,\thinspace \alpha}+g_{\alpha \rho,\thinspace \nu}-g_{\nu \alpha,\thinspace \rho} \right)
%\end{eqnarray}

To simplify the calculation, we assume that the trajectory of the test particle follows a circular orbit and is confined to the equatorial plane, where $\theta={\pi}/{2}$.
Here, we introduce the effective potential $V_{\mathrm{ep}}$ to analyze the stability of the test particle's orbit. From Eq. \eqref{equ:II.B-(33)}, we can derive:
\begin{eqnarray}
\dot{r}^{2}+V_{\mathrm{ep}}=E^2  ,
\end{eqnarray}
where the effective potential is
\begin{eqnarray}
V_{\mathrm{ep}}=E^2-\left(-1+\frac{E^2}{f\left(r\right)}-\frac{l^2}{r^2} \right)f\left(r\right)g\left(r\right)  .
\end{eqnarray}

To have a stable circular orbit in the equatorial plane, both the radial velocity and radial acceleration need to be zero simultaneously, which means $V_{\mathrm{ep}}=E^2$ and $\mathrm{d} V_{\mathrm{ep}}/{\mathrm{d}r}=0$. These conditions yield:
\begin{eqnarray}
{E}= \left(\frac{2f^{2}\left(r\right)}{2f\left(r\right)-f'\left(r\right)r}\right)^\frac{1}{2}, \quad
{l}= \left(\frac{r^{3}f'\left(r\right)}{2f\left(r\right)-f'\left(r\right)r}\right)^\frac{1}{2}.
\end{eqnarray}
Hence, the related four-velocity vectors can be recast as:
\begin{eqnarray}
u^{t} = \left(\frac{2}{2f\left(r\right)-f'\left(r\right)r}\right)^\frac{1}{2}, \quad
u^{\phi} = \left(\frac{f'\left(r\right)}{2rf\left(r\right)-f'\left(r\right)r^{2}}\right)^\frac{1}{2} .
\end{eqnarray}
By definition, we find that:
\begin{eqnarray}
\Omega \equiv \frac{\mathrm{d}\phi}{\mathrm{d}t} = \frac{u^{\phi}}{u^{t}} = \left(\frac{f'\left(r\right)}{2r}\right)^\frac{1}{2} ,
\label{equ:III.D-Omega1}
\end{eqnarray}
where $\Omega$ is the orbital angular velocity of the test particle.

Based on these results, we can form the parallel transport equations as follows:
\begin{eqnarray}
&&\frac{\mathrm{d} S^{t}}{\mathrm{d} \tau}+\frac{1}{2}\frac{f'\left(r\right)}{f\left(r\right)} S^{r} u^{t}=0,  \label{equ:III.D-St}\\[3mm]
&&\frac{\mathrm{d} S^{r}}{\mathrm{d} \tau}+\frac{1}{2}{f\left(r\right)g\left(r\right)}{f'\left(r\right)} S^{t} u^{t}-{r f\left(r\right)g\left(r\right)}S^{\phi} u^{\phi}=0,  \label{equ:III.D-Sr}\\[3mm]
&&\frac{\mathrm{d} S^{\theta}}{\mathrm{d} \tau}=0,  \label{equ:III.D-Stheta} \\[3mm]
&&\frac{\mathrm{d} S^{\phi}}{\mathrm{d} \tau}+\frac{1}{r} S^{r} u^{\phi}=0.  \label{equ:III.D-Sphi}
\end{eqnarray}

For convenience, we substitute the derivatives with respect to coordinate time $t$ for derivatives with respect to proper time $\tau$. It is worth noting that $\mathrm{d}\tau=\mathrm{d}t/u^{t}$, and this leads to the following expressions for the spin vectors:
\begin{eqnarray}
&&S^{t}\left(t\right) = - \frac{f'\left(r\right)}{2\omega}\sqrt{\frac{g\left(r\right)}{f\left(r\right)}}\sin\left(\omega t\right) , \\[3mm]
&&S^{r}\left(t\right)=\sqrt{f\left(r\right)g\left(r\right)}\cos\left(\omega t\right) , \\[3mm]
&&S^{\theta}\left(t\right) = 0 , \\[3mm]
&&S^{\phi}\left(t\right) = - \frac{\Omega}{r \omega}\sqrt{f\left(r\right)g\left(r\right)}\sin\left(\omega t\right) .
\end{eqnarray} 
where
\begin{eqnarray}
\omega = \Omega \left(f\left(r\right)g\left(r\right) - \frac{r}{2} g\left(r\right)f'\left(r\right) \right)^{\frac{1}{2}} ,
\label{equ:III.D-omega}
\end{eqnarray}
is the angular velocity of the spin vector.
Here we have assumed that the spin vector is radial directed at $t=0$, i.e. $S^{t}\left(0\right) = S^{\theta}\left(0\right) = S^{\phi}\left(0\right) = 0$. The coefficients can be gained by the conditions \eqref{equ:III.D-orth} and \eqref{equ:III.D-norm}.
Hence, we can utilize the discrepancy between $\Omega$ and $\omega$ to detect the geodetic effect.

Upon completing one orbit, where $\phi$ changes from 0 to $2\pi$, the corresponding coordinate time is $\delta{t}=2\pi/\Omega$. Consequently, the geodetic precession angle per revolution can be expressed as:
\begin{eqnarray}
\Delta{\Phi}_{\mathrm{geo}}=2\pi - \omega \delta{t} = 2\pi \left(1-\frac{\omega}{\Omega} \right).
\label{equ:III.D-Delta-phi}
\end{eqnarray}
To obtain experimental constraints, substituting Eq. \eqref{equ:III.D-omega} into Eq. \eqref{equ:III.D-Delta-phi} and expanding it as power series in terms of $M/r$ up to first order gives:
\begin{eqnarray}
\Delta{\Phi}_{\mathrm{geo}} \approx \frac{3\pi M}{r}\left(1+\frac{2r_0}{6-3r_0}\right) = \Delta{\Phi}_{\mathrm{GR}}\left(1+\frac{2r_0}{6-3r_0} \right) .
\end{eqnarray}
We can conclude that the geodetic precession angle increases with the LQG parameter when $r_0>0$. 

Detecting such phenomena can be quite challenging. Fortunately, the Gravity Probe B (GP-B) mission, equipped with four nearly perfect spherical gyroscopes and a star-tracking telescope, operates on a polar orbit around earth at an altitude of 642 km \cite{Everitt:2011hp}. GP-B measures the geodetic drift rate in the north-south direction, with the GR prediction being $-6066.1$ milliarcseconds per year.
The analysis of data from the four gyroscopes reveals a geodetic drift rate of $\Delta{\Phi}=\left(-6066.8 \pm 18.3 \right)$ milliarcseconds per year. This measurement provides bounds for $r_0$ and $\lambda$:
\begin{eqnarray}
0 <r_0 < 6.34\times 10^{-3} , \quad  0 < \lambda < 5.65\times 10^{-2}.
\end{eqnarray}

Additionally, lunar laser ranging (LLR) has proven to be one of the most powerful tools for rigorously testing GR theory with high level of precision \cite{muller2019lunar}.
The Earth-Moon system can be deemed as a gyroscope moving around the sun, the geodetic precession is manifested through the change of the lunar orbit which has already reach a level that can be observed by laser ranging. 
Thus, by measuring the lunar orbit within the Earth-Moon system's dynamic in the weak field of the sun, LLR provides a relative deviation of geodetic precession from GR value, yielding
$K_{\mathrm{gp}}= -0.0019 \pm 0.0064$. Based on this result, we therefore obtain upper limits for the parameter $r_0$ and $\lambda$ as:
\begin{eqnarray}
0 <r_0 < 1.34\times 10^{-2} , \quad  0 < \lambda < 8.24\times 10^{-2}.
\end{eqnarray}

%%%%%%%%%%%%%%%%%%%%%%%%%%%%%%%%%%%%%%%%%%
\section{Conclusion}\label{conclusion}

LQG is one of the candidates of quantum gravity theories. It offers a solution to the singularity problem, whether in cosmology or black hole physics. By introducing the concept of polymerization approach, spacetime is quantized, replacing singularities with a minimum area gap, which results in a spacelike transition surface to exterior space.
In this work, we investigate the classical tests of a LQG-corrected black hole  within an effective LQG framework. These tests encompass the light deflection, the Shapiro time delay, the perihelion precession, and the geodetic precession. Utilizing these classical observations, we calculate the impact of the LQG-corrected parameters, namely, $r_0$ and $\lambda$, and derive constraints on these parameters by incorporating the latest astronomical observations within the solar system. The corresponding results of this analysis are summarized in Table \ref{tab:summary}. 

We find it interesting that the LQG correction terms always put positive impacts on modified classical tests of GR, it may reflect the connection between the quantum scale effects and the macroscopic effects. As shown in Table \ref{tab:summary}, it is exciting to note that the Cassini solar conjunction experiment gives the most stringent upper bound on the parameter $r_0$ as $8.80\times 10^{-5}$. Note that the Doppler tracking method of the Cassini spacecraft also yields consistent results. In addition, the VLBI and MESSENGER ranging data also provide nice constraints as $0<r_0<2.0\times10^{-4}$ and $0<r_0<1.26\times10^{-4}$ respectively.

%%%----------------------------------------------------------------------------------------%%
\begin{table}
	\centering
	%\begin{tabular}{cccc}
	%  \setlength{\tabnotewidth}{1.0\columnwidth}
	%  \tablecols{4}
	%   \setlength{\tabcolsep}{2.8\tabcolsep}
	%\caption{C\lowercase{onstituents of the universe and  their behaviour: 
			%density evolution $\rho(a)$, scale factor $a(t)$}, H\lowercase{ubble parameter} $H(\lowercase{t})$.}
	\caption{\centering Summary of estimates for upper bounds of the quantum parameters $r_0$ and $\lambda$ in the covariant LQG black hole model from several astronomical observations.}
	\vspace{0.5em}
	\begin{tabular}{c|c|c|c}
		\toprule[1pt]
		\toprule[0.5pt]
		\midrule
		\quad {\rm Experiments/Observations \ } \quad  & \ \ $r_0$ \quad & $ \lambda $  \ & \quad Datasets \\
		\midrule
		\toprule[1pt]
		\midrule
		Light deflection &\quad $ 2.0\times 10^{-4} \quad$ & \quad $ 1.0\times 10^{-2} \quad$ & \  VLBI observation of quasars 	\\	
		\midrule
		\toprule[0.1pt]
		\midrule
		\multirow{2}{*}{Shapiro time delay} & \quad  $ 8.80\times 10^{-5} \quad$ & \quad $ 6.63\times 10^{-3} \quad$  & \ Cassini mission 	\\	
		& \quad  $ 4.0\times 10^{-5} \quad$ & \quad $ 4.47\times 10^{-3} \quad$  & \ Doppler tracking of Cassini 	\\
		\midrule
		\toprule[0.1pt]
		\midrule
		\multirow{3}{*}{Perihelion advance} & \quad $ 1.26\times 10^{-4} \quad$ & \quad $ 7.93\times 10^{-3} \quad$  &\ MESSENGER mission  \\
		& \quad $ 1.44\times 10^{-2} \quad$ & \quad $ 8.55\times 10^{-2} \quad$  &\ LAGEOS II satellites  \\
		& \quad $ 0.93 \quad$ & \quad $ 2.59 \quad$  & \ Observation of the S2-Sgr A$^{*}$ orbit \\
		\midrule
		\toprule[0.1pt]
		\midrule
		\multirow{2}{*}{Geodetic precession} & \quad $ 6.34\times 10^{-3} \quad$ & \quad $ 5.65\times 10^{-2} \quad$  &\ Gravity Probe B  \\
		& \quad $ 1.34\times 10^{-2} \quad$ & \quad $ 8.24\times 10^{-2} \quad$  &\
		Lunar Laser Ranging  \\
		\midrule
		\toprule[0.1pt]
		\midrule
		Strong equivalence principle test \ & \quad $ 5.93\times 10^{-4} \quad$ & \quad $ 1.72\times 10^{-2} \quad$  &\ Lunar Laser Ranging  \\
		\midrule
		\bottomrule[1pt]
	\end{tabular}
	\label{tab:summary}
	%\end{ruledtabular}
\end{table}
%%%----------------------------------------------------------------------------------------%%

We can estimate the scale of the quantum parameter. Reminder that the parameter here $r_0$ is rescaled by the mass of the central celestial object, rendering it a dimensionless quantity. The original dimensionful quantum parameter $r_0$ is a Planck scale quantity. Considering the central celestial object as the sun, we can easily estimate the dimensionless quantum parameter, finding that $r_0 \sim 10^{-38}$ and $\lambda \sim 10^{-20}$. Consequently, our theoretical estimation of the quantum parameter is well below the current observational bounds from solar system tests. It seems unlikely that such a value could be observationally tested in the solar system in the near future. Nevertheless, given the significant role of quantum gravity effects in the strong field regime, we anticipate the detection of quantum gravity effects in tests involving central celestial objects like BHs, especially in the observation of gravitational waves.

We also calculate the upper bounds for the polymerization parameter in the self-dual spacetime within LQG. The results are detailed in Appendix \ref{app-selfdual}. The upper bounds for the polymerization parameter $\delta$ in the self-dual black hole model, as constrained by solar system experiments, are roughly one order of magnitude higher than the upper bounds for the polymerization parameter $\lambda$ in our current model. Consequently, we can draw a similar conclusion to that of our current model.

We notice that the magnitude of upper bound on $r_0$ from S2 star orbit observations around $\mathrm{Sgr\ A^*}$ is much larger. As pointed out that in \cite{GRAVITY:2020gka}, it may be limited by experimental accuracy.
For the future optical observations, astrometric missions such as GAIA will push the accuracy to the microarcsecond level, thus the measure of light deflection due to the sun and the PPN parameter $\gamma$ will be hopefully reach the order of $10^{-6}$ or even better \cite{Crosta:2018nif, Gaia:2016zol}. Besides, the BepiColombo mission was launched on 20 October 2018 for the exploration of Mercury, which will give a higher precise of constraint on the value of $\gamma$ in the near future \cite{Will:2018mcj}. We look forward that with these missions the accuracy of constraints on quantum parameters in LQG will be improved.

\acknowledgments

We are especially grateful to Yun-Long Liu for helpful discussions and suggestions. This work is supported by National Key R$\&$D Program of China (No. 2020YFC2201400), the Natural Science Foundation of China under Grants No. 12375055.

%%%%%%%%%%%%%%%%%%%%%%%%%%%%%%%%%%%%%%%%%%
\section{Appendix}\label{Appendix}

\subsection{PPN APPROACH}

In Sec. \ref{constrain}, we have mentioned the PPN formalism in experimental data to analyze the LQG theoretical calculation results of classical tests and obtain the constraints on $r_{0}$. PPN formalism contains all post-Newtonian theory, which is a good approximation in the weak field regime, especially in the solar system, and slow motion \cite{Misner:1973prb, will2018theory}. 

The PPN limit of the Schwarzschild metric in isotropic coordinates provided by the standard form \cite{eddington1923mathematical}:
\begin{equation}
\mathrm{d}s^{2} =- A\left(r\right)\mathrm{d}t^{2}+B\left(r\right)\mathrm{d}r^{2}+r^{2}\left(\mathrm{d}\theta^2 +\sin^2\theta\thinspace \mathrm{d}\phi^2 \right) .
\end{equation}
In this metric, the coefficients $A\left(r\right)$ and $B\left(r\right)$ can be expanded as power series in terms of the small quantity $M/r$, as given by \cite{robertson1962space}:
\begin{eqnarray}
A\left(r\right) &=& 1- \frac{2M}{r}+2\left(\beta-\gamma\right)\frac{M^{2}}{r^{2}}+\cdots , \label{equ:III.A-Ar}   \\[3mm]
B\left(r\right) &=& 1+2\gamma \frac{M}{r}+\cdots . \label{equ:III.A-Br}  
\end{eqnarray}
The PPN parameter $\gamma$ provides a rough description of the amount of space-curvature produced by unit rest mass, while $\beta$ gives a rough indication of the nonlinearity in the superposition of gravity, as described in \cite{will2015gravity}. According to Einstein's theory, both parameters are predicted to have strict values of $\gamma=\beta=1$.

Expanding the coefficients of $\mathrm{d}t$ and $\mathrm{d}r$ components in metric \eqref{equ:II.A-(1)} yield:
\begin{eqnarray}
f\left(r\right) &=& 1- \frac{2 M}{r}, \label{equ:III.A-fr}   \\[3mm]
\frac{1}{g\left(r\right)f\left(r\right)} &=& 1 + 2\left(\frac{2}{2-r_0}\right)\frac{M}{r} + \mathcal{O}\left(\frac{M}{r}\right)^2 .  \label{equ:III.A-1/grfr}  
\end{eqnarray}
By comparison, we see that $\beta=\gamma=2/\left(2-{r_0}\right)$ matches the related expression in Sec. \ref{constrain}. This method allows us to test whether the calculations are consistent with the situation under the weak field limit.

For instance, the PPN correction factor in the perihelion advance is combined with $\beta$ and $\gamma$ forms \cite{Will:2014kxa, will2018theory, park2017precession}:
\begin{eqnarray}
	\Delta{\phi} = \frac{6 \pi M_{\odot}}{a \left(1-e^2 \right)}\left(\frac{2-\beta+2\gamma}{3} \right) = \frac{6 \pi M_{\odot}}{a \left(1-e^2 \right)}\left(\frac{6-2r_0}{6-3r_0} \right).
\end{eqnarray}
which is equivalent to the LQG correction in Eq. \eqref{equ:III.C-delta phi0}.

Another application involves the LLR strong equivalence principle (SEP) test \cite{muller2019lunar}. In this context, the PPN coefficient $\eta=4\beta-\gamma-3$, characterizes the strength of violations of the Einstein equivalence principle (EEP). This test provides upper constraints on the parameters $r_0$ and $\lambda$, yielding:
\begin{eqnarray}
0 <r_0 < 5.93\times 10^{-4} , \quad  0 < \lambda  < 1.72\times 10^{-2}.
\end{eqnarray}

\subsection{UPPER BOUNDS OF THE POLYMERIZATION PARAMETER IN THE SELF-DUAL SPACETIME IN LQG}\label{app-selfdual}

Currently, numerous LQG-corrected BH models have been proposed \cite{Ashtekar:2005qt,Modesto:2005zm,Modesto:2008im,Modesto:2009ve,Campiglia:2007pr,Bojowald:2016itl,Boehmer:2007ket,Chiou:2008nm,Chiou:2008eg,Joe:2014tca}. Most of these BHs are characterized by the polymerization parameter from LQG, denoted as $\lambda$ in this paper. We anticipated that the fundamental polymerization parameter shares the same scale. Consequently, it is intriguing to compare the constraints on the polymerization parameter from solar system tests among different LQG-corrected black hole models. This appendix provides a comparison of the constraints on the polymerization parameter from solar system tests between our current model and the self-dual spacetime in LQG.

\begin{table}
	\centering
	%\begin{tabular}{cccc}
	%  \setlength{\tabnotewidth}{1.0\columnwidth}
	%  \tablecols{4}
	%   \setlength{\tabcolsep}{2.8\tabcolsep}
	%\caption{C\lowercase{onstituents of the universe and  their behaviour: 
			%density evolution $\rho(a)$, scale factor $a(t)$}, H\lowercase{ubble parameter} $H(\lowercase{t})$.}
	\caption{\centering Summary of upper bound estimates for the polymerization function $P$ and the parameter $\delta$ in the self-dual spacetime in LQG from various observations.}
	\vspace{0.5em}
	\begin{tabular}{c|c|c|c}
		\toprule[1pt]
		\toprule[0.5pt]
		\midrule
		Experiments   &   $P$    &    $\delta$    &   Datasets \\
		\midrule
		\toprule[1pt]
		\midrule
		Light deflection & \quad $ 2.50\times 10^{-5} \quad$ & \quad $ 4.21\times 10^{-2} \quad$ & \  VLBI observation of quasars 	\\	
		\midrule
		\toprule[0.1pt]
		\midrule
		Shapiro time delay & \quad  $ 5.0\times 10^{-6} \quad$ & \quad $ 1.88\times 10^{-2} \quad$  & \ Doppler tracking of Cassini 	\\
		\midrule
		\toprule[0.1pt]
		\midrule
		\multirow{3}{*}{Perihelion advance} & \quad $ 7.85\times 10^{-6} \quad$ & \quad $ 2.36\times 10^{-2} \quad$  &\ MESSENGER mission  \\ 
		& \quad $ 9.05\times 10^{-4} \quad$ & \quad $ 2.54\times 10^{-1} \quad$  &\ LAGEOS II satellites  \\
		& \quad $ 8.63\times 10^{-2} \quad$ & \quad $ 2.71 \quad$  & \ Observation of the S2-Sgr A$^{*}$ orbit \\
		\midrule
		\toprule[0.1pt]
		\midrule
		\multirow{2}{*}{Geodetic precession} & \quad $ 7.93\times 10^{-4} \quad$ & \quad $ 2.37\times 10^{-1} \quad$  &\ Gravity Probe B  \\
		& \quad $ 1.68\times 10^{-3} \quad$ & \quad $ 3.46\times 10^{-1} \quad$  &\ Lunar Laser Ranging  \\
		\midrule
		\bottomrule[1pt]
	\end{tabular}
	\label{tab:zhu}
	%\end{ruledtabular}
\end{table}
%%%----------------------------------------------------------------------------------------%%

In Ref. \cite{Zhu:2020tcf}, the authors have studied the observational tests of the self-dual spacetime in LQG within the solar system context. However, it is crucial to emphasize that, to recover the Newtonian limit, we establish a relationship between the effective gravitational parameter and Newton's gravitational constant as $G_\mathrm{N}=G\left(1-P\right)^2/\left(1+P\right)^2$ \cite{Liu:2023vfh, Yan:2022fkr}. Through this transformation, we reevaluate the constraints on the polymerization parameter in the self-dual spacetime in LQG. The results are presented in Table \ref{tab:zhu}\footnote{We adopt the convention from Refs. \cite{Zhu:2020tcf,Liu:2023vfh, Yan:2022fkr}, representing the polymerization parameter as $\delta$.}.

Comparing Table \ref{tab:zhu} with Table \ref{tab:summary}, we observe that the upper bounds of the polymerization parameter $\delta$ in the self-dual black hole model, constrained by solar system experiments, are approximately one order of magnitude larger than the upper bounds of the polymerization parameter $\lambda$ in our present model. Similar to the discussion in this paper, we can find that the theoretical estimation of the polymerization parameter falls well below the current observational bounds from solar system tests. 
We anticipate that increasingly precise experiments will provide a more thorough elucidation of the shared characteristics among LQG black holes.

\bibliographystyle{utphys}
\bibliography{Ref}
\end{document}